\documentclass[journal]{IEEEtran}
\IEEEoverridecommandlockouts
\usepackage{cite}
\usepackage{makecell}
\usepackage[utf8x]{inputenc}
\usepackage{graphicx}
\usepackage{subfigure}
\usepackage{array}
\usepackage{booktabs}
\usepackage{url}
\usepackage[misc]{ifsym}
\usepackage[dvips]{color}
\usepackage{mathrsfs}
\usepackage[linesnumbered,ruled,vlined]{algorithm2e}
\usepackage{amsmath}
\usepackage{nccmath}
\usepackage{amsfonts,amssymb,bm}
\usepackage{amsthm}
\usepackage[dvipsnames]{xcolor}
\usepackage[noend]{algpseudocode}
\usepackage{overpic}
\usepackage{multirow}
\usepackage{algorithm2e}
\usepackage{algpseudocode}

\SetCommentSty{mycommfont}
\SetKwInput{KwInput}{Input}
\SetKwInput{KwOutput}{Output}

\begin{document}
\title{Contact-Free Multi-Target Tracking Using Distributed Massive MIMO-OFDM Communication System: Prototype and Analysis}
\author{Chenglong~Li,~\IEEEmembership{Member,~IEEE,}
		Sibren~De Bast,~\IEEEmembership{Student Member,~IEEE,}
		Yang~Miao,~\IEEEmembership{Member,~IEEE,}
		Emmeric~Tanghe,~\IEEEmembership{Member,~IEEE,}
		Sofie~Pollin,~\IEEEmembership{Senior~Member,~IEEE}
		and Wout~Joseph,~\IEEEmembership{Senior~Member,~IEEE}
\thanks{This work is supported in part by the Excellence of Science (EOS) project MUlti-SErvice WIreless NETworks (MUSE-WINET), by the Research Foundation Flanders (FWO) strategic basic (SB) PhD fellowship under Grant 1SA1619N, and by the FWO project under Grant G098020N.}
\thanks{C. Li, E. Tanghe, and W. Joseph are with the WAVES, Department of Information Technology, Ghent University-imec, 9052 Ghent, Belgium (e-mail: chenglong.li@ugent.be).}
\thanks{S. De Bast and S. Pollin are with Wavecore, Department of Electrical Engineering, KU Leuven, 3001 Leuven, Belgium. S. Pollin is also with imec, 3001 Leuven, Belgium}
\thanks{Y. Miao is with the Faculty of Electrical Engineering, Mathematics and Computer Science, University of Twente, 7522 Enschede, The Netherlands, and also with Wavecore, Department of Electrical Engineering, KU Leuven, 3001 Leuven, Belgium.}
}

\markboth{}%
{Shell \MakeLowercase{\textit{et al.}}: Bare Demo of IEEEtran.cls for IEEE Journals}

\maketitle
\begin{abstract}
Wireless-based human activity recognition has become an essential technology that enables contact-free human-machine and human-environment interactions. In this paper, we consider contact-free multi-target tracking (MTT) based on available communication systems. A radar-like prototype is built upon a sub-6 GHz distributed massive multiple-input and multiple-output (MIMO) orthogonal frequency-division multiplexing (OFDM) communication system. Specifically, the raw channel state information (CSI) is calibrated in the frequency and antenna domain before being used for tracking. Then the targeted CSIs reflected or scattered from the moving pedestrians are extracted. To evade the complex association problem of distributed massive MIMO-based MTT, we propose to use a complex Bayesian compressive sensing (CBCS) algorithm to estimate the targets' locations based on the extracted target-of-interest CSI signal directly. The estimated locations from CBCS are fed to a Gaussian mixture probability hypothesis density (GM-PHD) filter for tracking. A multi-pedestrian tracking experiment is conducted in a room with a size of 6.5 m$\times$10 m to evaluate the performance of the proposed algorithm. According to experimental results, we achieve 75th and 95th percentile accuracy of 12.7 cm and 18.2 cm for single-person tracking and 28.9 cm and 45.7 cm for multi-person tracking, respectively. Furthermore, the proposed algorithm achieves tracking purposes in real-time, which is promising for practical MTT use cases.

\end{abstract}


\IEEEpeerreviewmaketitle

\section{Introduction}
\IEEEPARstart{H}{uman} activity recognition (HAR) using wireless devices has attracted considerable attention due to the advances in Internet-of-Things technology and radio frequency (RF) hardware circuits in the past decade \cite{Zijuan2019}. It enables a wide variety of potential applications, such as elderly care, behavior analysis, virtual reality, location-based services (LBS), etc. Generally, RF-based HAR techniques can be classified into two categories, namely, wearable device-based and device-free (\textit{a.k.a.}, contact-free or passive) solutions. Specifically, contact-free human sensing has become a more active research area as there is no need for user-attached sensors, which provides convenience and feasibility to specific use cases, such as infectious patient monitoring. Moreover, RF-based techniques are more privacy-preserving compared to image-based solutions. To date, a batch of academic and industrial efforts have been paid to promote human sensing techniques based on different standards, including radio frequency identification (RFID), Bluetooth low energy (BLE), ultra-wideband (UWB), Wi-Fi, millimeter wave, etc. 
\par
Among them, many efforts have been devoted to Wi-Fi-based human sensing because of the benefit of ubiquitous deployment. Wang \textit{et al.} \cite{Wei2017} proposed a channel state information (CSI)-based human activity monitoring algorithm by quantifying how the speed of human motion is related to the CSI power variations. Xu \textit{et al.} \cite{Qinyi2019} proposed to exploit the temporal CSI series to characterize the dynamic indoor motion event and realized high accuracy for real-time monitoring. Wi-Fi-based CSI signals have also been widely used for the single-target contact-free tracking \cite{Xiang2017,widar2,Kai2022}, where the person is located via the geometrical metrics, such as angle-of-arrival (AoA), Doppler, time-of-flight (ToF), etc. Rather than single-target tracking, Karanam \textit{et al.} \cite{Karanam2019} extended the work to multiple-person localization leveraging the AoA and Doppler estimates of the magnitude of CSI. Then, a particle filter together with joint probabilistic data association (JPDA) was used for tracking. Venkatnarayan \textit{et al.} \cite{Venkatnarayan2020} proposed to add horizontal and vertical polarization to perform multi-person tracking and achieved higher accuracy than the case without polarization diversity. However, this has a more rigorous demand for transceiver deployment. Besides Wi-Fi, UWB and millimeter-wave radios have also been widely used for contact-free human sensing due to the high-ranging resolution, for instance, vital sign monitoring \cite{Tianyue2020,Fengyu2021}, pedestrian tracking \cite{Chenshu2020,Chenglong2022}, etc. 
\par
As the standardization of fifth-generation (5G) wireless communication gradually solidifies, people are envisioning what the next-generation network will be. One of the great ambitions is to enable the wireless network to ``see" or ``understand" the physical world via sensing the targets or surroundings. Towards this end, the so-called integrated sensing and communication (ISAC) \cite{Carlos2021,Liu2022} has been proposed which integrates the sensing functionality into wireless communication networks allowing the reuse of allocated spectrum, hardware, and even signaling resources. Rather than redesigning the whole hardware architecture, one of the feasible and promising schemes is to re-utilize the existing and pervasive RF signals, \textit{e.g.}, Wi-Fi, Long-Term Evolution (LTE), and 5G, for the sensing objectives \cite{Daqing2022}. As a milestone of the 5G wireless communication, the massive multiple-input multiple-output (MIMO) technique not only improves communication in terms of channel capacity and spectral efficiency but also has the potential for accurate LBS applications resulting from the high spatial resolution \cite{MaMIMOLoc2022,bjornson2019}. 
\par
Despite the increasing discussions about enabling sensing via communication systems or the communication/radar co-design, little work is done on the experiment-based massive MIMO for ISAC except for our preliminary works in \cite{Sakhnini2022,Li2022b} for single-target sensing. To this end, in this paper, we prototype a sub-6 GHz distributed massive MIMO communication system and investigate its radar-like functionality for human sensing. The established massive MIMO system adopts a standard cellular signal bandwidth and the orthogonal frequency-division multiplexing (OFDM) waveform, which is compatible with the available Wi-Fi, LTE, and sub-6 GHz 5G communication systems. Inspired by the user-centric concept of cell-free (or distributed) massive MIMO \cite{interdonato2019}, which aims to surround the users with a large number of base station antennas, we implement a distributed massive MIMO radar-like system with a large antenna array separated in an indoor environment. Based on the prototype, we conduct the contact-free multiple pedestrians tracking experiment and analysis. The major contributions of this paper are as follows,
\begin{enumerate}
\item We demonstrate a radar-like distributed massive MIMO system using an OFDM communication testbed. A standard cellular signal bandwidth, \textit{i.e.}, 18 MHz, is adopted. According to the authors' best knowledge, it is the first bandwidth-limited distributed massive MIMO prototype for contact-free human sensing.
\item Based on the established prototype, we investigate the multi-target tracking (MTT) problem and propose a contact-free pedestrian tracking framework based on CSI data directly. Specifically, a serial interference cancellation and reconstruction (SICAR) algorithm is proposed for the targeted CSI extraction. The complex Bayesian compressive sensing (CBCS) and Gaussian mixture probability hypothesis density (GM-PHD) filter are introduced for positioning and tracking purposes. The proposed algorithm effectively evades complex association problems by exploiting the ideas of space sparsity and random finite set.
\item The MTT performance is evaluated comprehensively concerning the impact of bandwidth, the number of antennas, antenna deployment, the number of targets, etc. Besides, within the proposed MTT framework, each algorithm block is assembled organically, which helps to realize (quasi-) real-time MTT. 
\item Furthermore, to evaluate contact-free MTT, we propose an active tracking algorithm built upon the idea of the synthetic aperture. A centimeter-level (even millimeter-level) accuracy is guaranteed based on the proposed active tracking algorithm, which is feasible as the benchmark of the contact-free MTT. The idea of the proposed method can also be generalized to other device-based tracking technologies, such as phase-based RFID tracking, phase-based BLE tracking, CSI-based Wi-Fi tracking, etc.
\end{enumerate}
\par
The remainder of this paper is organized as follows. Section~\ref{sec:ExpSys} introduces the established massive MIMO prototype, experimental campaign, and system overview. The practical challenges for MTT are also discussed in this section. Section~\ref{sec:Prepara} presents the CSI calibration and benchmark design for contact-free tracking. Section~\ref{sec:AlgDesign} includes the ToI extraction, instantaneous locations estimation, and MTT algorithms design. The experimental results and analysis are presented in Section~\ref{sec:PerfEva}. Section~\ref{sec:conclusion} concludes this paper.

\section{Experiment and System Overview}
\label{sec:ExpSys}

\begin{figure}[t]
\centering
\includegraphics[width=0.485\textwidth]{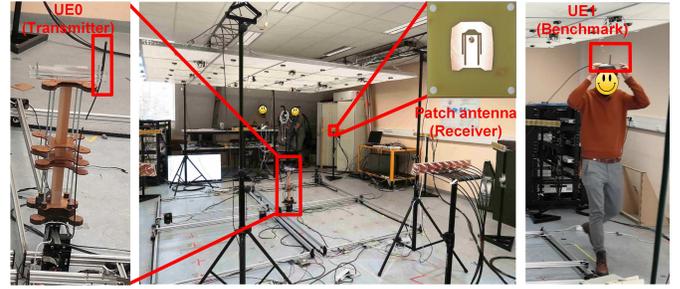}
\caption{Measurement setups of the distributed massive MIMO radar-like system \cite{Li2022b}. The uplink scenario is considered: user equipment (UE0, transmitter) is deployed in the center of the targeted area and the receiver consists of 64 patch antennas. UE1 on the top of the participant's head is used for active tracking as a benchmark.}
\label{fig:SetupPics}
\end{figure}

\begin{table}[t]
\centering
\caption{Distributed massive MIMO-OFDM communication system parameters setting}
\begin{tabular}{l|c}
\hline
\textbf{System parameters} & \textbf{Value}          \\ \hline\hline
\textbf{Transmitter power} & 15 dBm					\\ \hline
\textbf{Sampling rate}     & 100 Hz  				\\ \hline
\textbf{Center frequency}  & 2.61 GHz 				\\ \hline
\textbf{Wavelength}        & 11.49 cm				\\ \hline
\textbf{Bandwidth}         & 18 MHz  				\\ \hline
\textbf{\# of Sub-carrier}  & 100     				\\ \hline
\textbf{\# of antenna}     & 64 	  				    \\ \hline
\textbf{Modulation}        & QPSK    				\\ \hline
\textbf{Patch antenna size}& 7 cm$\times$7 cm  		\\ \hline
\textbf{BS antenna altitude}& 1.205 m    			    \\ \hline
\textbf{UE antenna altitude}& 0.8 m      			    \\ \hline
\textbf{Size of experimental space}& 6.5 m$\times$10 m  \\ \hline
\end{tabular}
\label{table:Paras}
\end{table}

\begin{figure*}[t]
\centering
\includegraphics[width=0.8\textwidth]{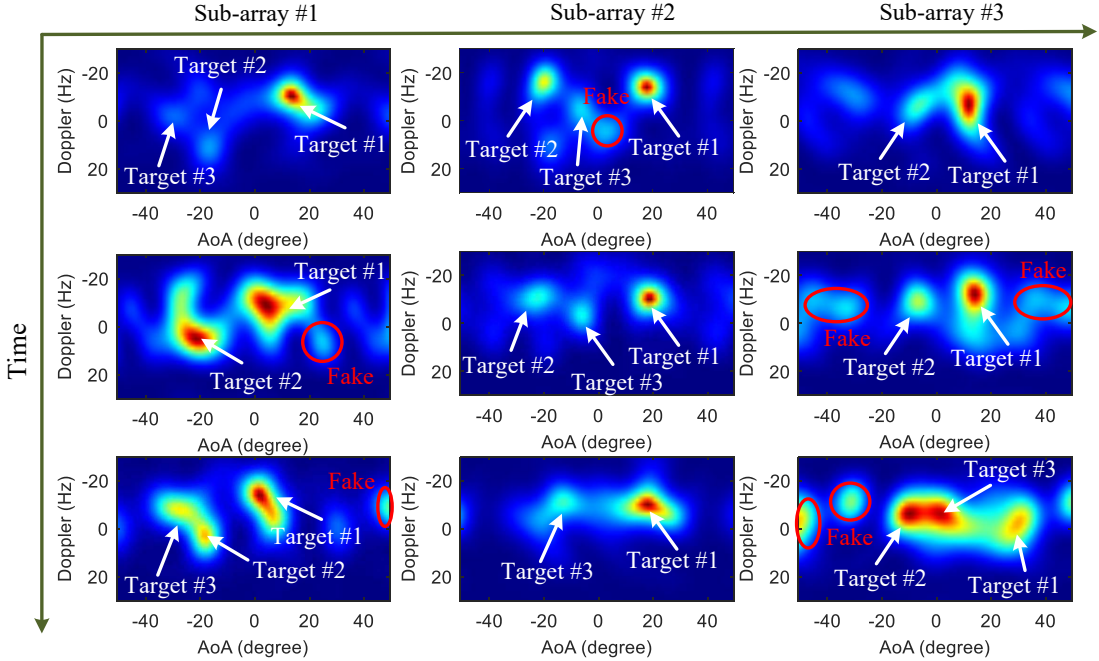}
\caption{Angular-Doppler profiles at three sub-antenna arrays based on a 2-D MUSIC algorithm. There are three moving pedestrians inside the targeted area. The red circles/ellipses indicate the false alarms. The horizontal axis denotes the sub-array index and the longitudinal axis indicates the time evolving.}
\label{fig:AoA_DPL_MUSIC}
\end{figure*}

\subsection{Prototype and Experiment}
\label{sec:campaign}
We establish a radar-like prototype based on the massive MIMO communication testbed at KU Leuven ESAT-WAVECORE. The Massive MIMO system is built on time division duplex (TDD)-based OFDM signaling. In the system, the massive number of the base station (BS) antennas are deployed as 8 distributed uniform linear arrays (ULA), each composing 1$\times$8 patch antennas, as shown in Fig.~\ref{fig:SetupPics}. For the investigation in this paper, only the uplink transmission is considered. Namely, the user equipment (UE) acts as the transmitter and the BS is the receiver. During the channel measurement, the BS is under the control of the NI LabVIEW Communications MIMO Application Framework \cite{NIreport2019}. This software framework runs on 32 Universal Software Radio Peripherals (URSPs) simultaneously, each controlling two BS antennas. For the massive MIMO system, there are 1200 subcarriers (with 15 kHz spacing for 18 MHz bandwidth) in use, which are divided into 100 resource blocks of 12 subcarriers to support up to 12 users. Each UE transmits a pilot on a different subcarrier in such a single resource block. The pilot tone of each UE consists of 100 sub-carriers, which are evenly spaced in a frequency band of 18 MHz. The BS (all 64 antennas) receives the orthogonal pilots sent by the UEs simultaneously and distinguishes between the different UEs by frequency interleaving of the subcarriers. In this way, the channel is captured between the UEs and the 64 BS antennas for 100 subcarriers. The system parameters setting are presented in Table~\ref{table:Paras}. 
\par
For the contact-free MTT experiment, the communication links among the transmitter (\textit{i.e.}, UE0) and the distributed BSs form a sensing zone, where multiple moving pedestrians (up to three in this paper) can be localized and tracked via characterizing the reflected/scattered signals on the human body. We have predefined the track templates for the participants to move. However, it should be noted that the participants cannot exactly follow the template tracks due to distinct individual movements. To handle this problem, we asked each participant to carry an additional transmitter (\textit{i.e.}, UE1, UE2, or UE3) for active tracking to determine a ``ground truth". To avoid the possible body shadowing effect, we placed the antenna on the top of the participant's head, as shown in Fig.~\ref{fig:SetupPics}. Therefore, the measured CSI of a single transmission can be represented by the complex matrix, given as follows,
\begin{equation}\label{eq:CSI_H}
\mathbf{H}_{\rm{CSI}} = \left\lbrace H_{n_r,n_f,n_{\rm{UE}}}\right\rbrace\in\mathbb{C}^{64\times100\times4},
\end{equation}where $n_r\in\lbrace1,2,\cdots,64\rbrace$, $n_f\in\lbrace1,2,\cdots,100\rbrace$, $n_{\rm{UE}}\in\lbrace0,1,2,3\rbrace$ represent the index of antenna, sub-carrier, and UE, respectively. In this work, we use the CSI of UE0 for contact-free tracking and the CSI of UE1, UE2, and UE3 for active tracking. The active tracking results are regarded as the benchmark for contact-free tracking. The feasibility of benchmarking will be validated in Section~\ref{sec:benchmark}.

\begin{figure*}[t]
\centering
\includegraphics[width=0.92\textwidth]{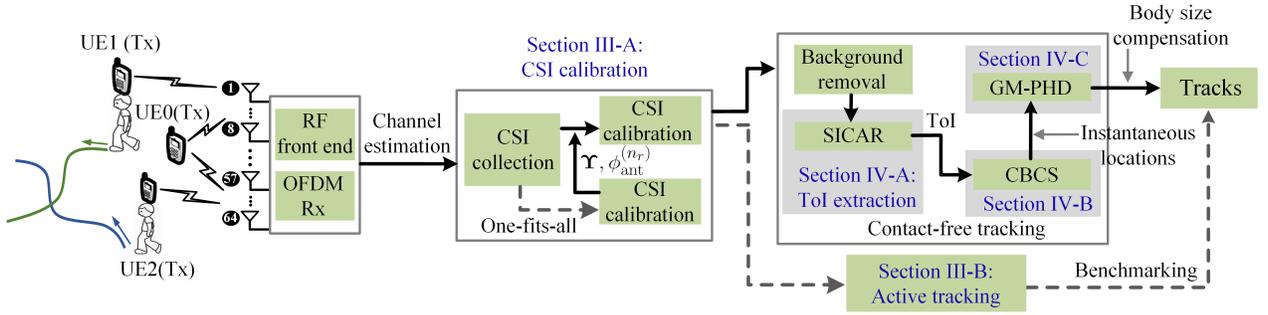}
\caption{Overall architecture of the proposed massive MIMO-based MTT system together with the active tracking as a benchmark. SICAR: serial interference cancellation and reconstruction. ToI: target-of-interest. CBCS: complex Bayesian compressive sensing. GM-PHD: Gaussian mixture probability hypothesis density.}
\label{fig:OverallSys}
\end{figure*}

\subsection{Practical Challenges}
\label{sec:challenges}
In recent years, there are increasing work on CSI-based contact-free tracking. However, most of them are focusing on single-target tracking (\textit{e.g.}, one walking person, a waving limb or finger) \cite{Xiang2017,widar2,Kai2022,Lei2021}. The common idea behind these pioneer works is to extract the location- or velocity-related parameters from the CSI. For example, in \cite{widar2}, the authors adopted a space-alternating generalized expectation-maximization (SAGE) algorithm to estimate the multipath components (MPCs), \textit{i.e.}, amplitude, AoA, Doppler, and ToF. Then the association of MPCs reflected by the body along the time evolving was regarded as an assignment problem and solved via binary optimization and the Hungarian algorithm. As a single person walking in the scenario is \textit{a priori}, the MPCs track with the highest power was selected. So for a single moving target, associating along the time domain is required for the tracking purpose. We name it \textbf{evolving association}. According to the authors' best knowledge, there are also a few works on CSI-based multi-person tracking \cite{Karanam2019,Venkatnarayan2020,Storrer2021}. Besides the location- or velocity-related parameters extraction as in single-person tracking, we need to assign the estimated MPCs to the corresponding targets as we have no idea of which targets that MPCs are generated from. So there is an additional \textbf{target association} for MTT. 
\par
However, this is not the end of the challenges for our distributed massive MIMO-based MTT. For the contact-free tracking research mentioned above, most of them just used a single transceiver link or two links for accuracy enhancement. But in the case of distributed massive MIMO, there are a large number of propagation links. The MPCs of the multiple targets along the large antenna array are also varied for a given time slot, which means we also need to assign the massive and cluttered MPCs of all the links to the corresponding target, namely, \textbf{link association}. Even though many algorithms can be used to solve the assignment problem, such as JPDA \cite{Karanam2019,Storrer2021}, multiple hypothesis tracking (MHT) \cite{Blackman2004}, binary optimization \cite{widar2}, etc., joint association in these three domains is challenging and complex in computation, which is not feasible for many practical applications considering the computing resource and run-time issue.
\par
Besides complex association, there are also like varied reflected power from the multiple targets, false alarms, and missed detection, challenging the distributed massive MIMO-based MTT problem. As presented in Section~\ref{sec:campaign}, we have deployed the distributed massive MIMO as 8 short sub-antenna arrays (ULA). We presume the planar wavefront assumption is satisfied for each short ULA. We adopt the consecutive CSI measurements within a given time window to estimate the Doppler. We set the time window as 0.1 s in this work, which is also widely adopted for person tracking \cite{widar2,Zhongqin2022}. Then a two-dimensional (2-D) multiple signal classification (MUSIC) algorithm together with spatial-temporal smoothing \cite{Karanam2019} is implemented to generate the angular-Doppler spectrum. Fig.~\ref{fig:AoA_DPL_MUSIC} shows the angular-Doppler profiles of three consecutive time windows in the case of three pedestrians. Three phenomenons can be observed from Fig.~\ref{fig:AoA_DPL_MUSIC}, which bring in even greater challenges to association problems of distributed massive MIMO-based MTT:
\begin{enumerate}
\item The reflected/scattered power from each target may be greatly varied. The major reasons are two-fold: First, the different distances from the pedestrian to the transceivers. Second, the different radar cross-sections (RCS) due to individual differences, body orientations, torso or limbs scattering, etc.
\item False alarms or ghost targets are prevalent. The possible reasons are three-fold. The first one is the residual components due to not perfect background mitigation. Second, the walking pedestrians may partly block the reflections/scattering from the static objects causing the signal from the static objects to fluctuate. Third, the second- or higher-order signal bounces within human bodies or between the bodies and the surroundings.  
\item Missed detection frequently happens. At some specific time slots or sub-antenna arrays, one or more targets cannot be detected, or several targets are merged from the observations. But it is possible to detect/separate them at other time slots or by other sub-antenna arrays.
\end{enumerate}

\subsection{Radar-Like System Overview}
\label{sec:sys_overview}
Different from conventional contact-free tracking solutions that exploit the estimated angle, distance, and Doppler estimates, in this paper, we propose to achieve contact-free MTT based on CSI data directly. In the proposed framework, the CSI measured by each antenna element of the massive MIMO system is regarded as a whole for positioning and tracking, which is similar to the idea of a synthetic aperture. This kind of processing has two merits. First, it can decrease the probability of missed detection. Because if the target is not detected by part of the antennas, it is likely the target can be detected by other antennas due to the distributed deployment. Second, we treat the massive antenna array as a whole and estimate the locations of the targets, which avoids the \textbf{link association} problem in the MPCs-based solutions mentioned above. Given the fact that the targets' states are sparse relative to the whole targeting space. So the locations can be estimated from the CSI matrix of the massive antennas via a compressive sensing algorithm. In this section, we summarize the whole procedures of the distributed massive MIMO prototype, including the CSI calibration, contact-free MTT algorithm design, and the benchmark for evaluation.
\par
The architecture of the established radar-like system is given in Fig.~\ref{fig:OverallSys}. For a given scenario, several UEs are deployed, of which one UE (\textit{i.e.}, UE0) acts as a transmitter and is used for contact-free tracking, whereas the other UEs (UE1, UE2, and UE3) are used for benchmarking which are not necessary for practical MTT tasks. It is noteworthy that for a real-world deployment, UE0 (transmitter) can be part of the infrastructure (like BS) or a cooperative user. For the latter case, the location of the user should be obtained first via active localization. 
\par
In the established prototype, the BS receives the signals transmitted by all UEs, conducts the channel estimation, and obtains the CSI data. The procedures for achieving the contact-free MTT are as follows:
\begin{enumerate}
\item \textbf{CSI calibration}: Before utilizing CSI for contact-free pedestrian tracking and benchmarking, CSI calibration is required along frequency and antenna domains \cite{MaMIMOLoc2022}, which will be introduced in Section~\ref{sec:calibration}. In the CSI calibration, we can obtain a set of calibration parameters that are fixed for a given system setting (frequency, bandwidth, antenna, etc.). So we can use these parameters for the future CSI calibration directly without the need for redundant processing as in Section~\ref{sec:calibration}, namely, one-fits-all CSI calibration.
\item \textbf{Target-of-interest (ToI) extraction}: First, we conduct background removal by subtracting the moving average CSI  \cite{Sakhnini2022,Storrer2021,Berger2010}. Then we propose a serial interference cancellation and reconstruction algorithm to extract the CSI related to the multiple targets, namely ToI, which will be discussed in Section~\ref{sec:ToI}.
\item \textbf{Instantaneous locations estimation}: We regard the CSI of the massive antenna array as a whole, and estimate the locations of the multiple targets directly. In Section~\ref{sec:CBCS}, we will introduce the CBCS algorithm to estimate the instantaneous locations based on the extracted ToI. This can be achieved due to the fact that the locations of the targets are sparse relative to the entire monitoring area.
\item \textbf{Multi-target tracking}: The estimated positions via the CBCS algorithm may be cluttered. In Section~\ref{sec:MTT}, the GM-PHD filter is adopted to track multiple targets simultaneously without the exact prior knowledge of the number of targets. GM-PHD filter is iterated within a Kalman recursion which ensures a real-time MTT.
\item \textbf{Benchmark and evaluation}: In Section~\ref{sec:benchmark}, we propose to use active tracking results as the benchmark of the contact-free MTT. Stemming from the concept of synthetic aperture, the CSI of all distributed antennas is synthesized to generate the location likelihood. A particle filter is introduced to track the varying likelihood when the target is moving. The proposed tracking algorithm can guarantee centimeter-level, even millimeter-level accuracy, which is feasible for benchmarking. Furthermore, the body size compensation is considered for the accuracy evaluation, which will be explained in Section~\ref{sec:SingleEva}.
\end{enumerate}

\begin{figure}[t]
\centering
\includegraphics[width=0.47\textwidth]{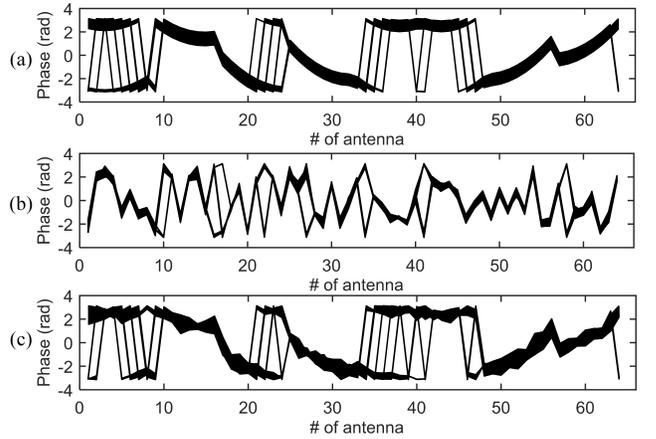}
\caption{The phases of 100 sub-carriers at 64 distributed antenna elements: (a) The expected phases based on the phase rotation of the LoS component. (b) Phases of raw CSI. (c) Phases of the calibrated CSI.}
\label{fig:CSI_calibration}
\end{figure}

\section{Preliminary Preparations}
\label{sec:Prepara}
\subsection{CSI Calibration}
\label{sec:calibration}
Due to the imperfect synchronization and hardware signal processing, the measured raw CSI suffers from various frequency-dependent phase errors, including the sampling frequency offset (SFO) $\phi_{\rm{SFO}}$, symbol timing offsets (STO) $\phi_{\rm{STO}}$, in-phase quadrature-phase (I-Q) imbalance  $\phi_{\rm{IQ}}$, and the constant carrier phase offsets (CPO) $\phi_{\rm{CPO}}$ \cite{MaMIMOLoc2022,ZhuSplicer2018}, which makes the CSI intractable for localization and tracking purposes. To this end, we need to calibrate the CSI before the radar experiment. According to \cite{MaMIMOLoc2022}, the mentioned phase offsets can be effectively mitigated via a nonlinear regression along the sub-carriers ($n_f=1,\cdots,N_f=100$), given by
\begin{equation}\label{eq:NLregression}
\mathop {\arg\min}\limits_{{\bf{\Upsilon}}}\!\sum\limits_{n_f}\left(\Delta{\Phi}_{n_f}\!-\!\phi_{\rm{IQ}}^{(n_f)}\!-\!n_f\phi_{\rm{SFO/STO}}\!-\!\phi_{\rm{CPO}}\right)^2,
\end{equation}where ${\bf{\Upsilon}}=[\varepsilon_g,\varepsilon_t,\varepsilon_p,\phi_{\rm{SFO/STO}},\phi_{\rm{CPO}}]$. $\Delta\Phi$ is the residual phase after removing the calculated LoS based on ground truth. The phase error due to the I-Q imbalance is denoted as \cite{ZhuSplicer2018},
\begin{equation}\label{eq:IQ}
\phi_{\rm{IQ}}^{(n_f)}=\arctan{\left(\varepsilon_g\frac{\sin{(n_f\varepsilon_t+\varepsilon_p)}}{\cos{(n_f\varepsilon_t)}}\right)},
\end{equation}where $\varepsilon_g$, $\varepsilon_p$, and $\varepsilon_t$ represent the gain mismatch, phase mismatch, and unknown time offset, respectively. Moreover, due to the heterogeneity of hardware, there are also constant phase offsets among each element of the large antenna array. The phase offsets at $n_r$-th antenna element ${\phi_{\rm{ant}}^{(n_r)}}$ can be estimated in complex domain via,
\begin{equation}\label{eq:ant}
\!\mathop {\arg\min}\limits_{{\phi_{\rm{ant}}^{(n_r)}}}\!\left\vert\sum\limits_{n_{\rm{RP}}}\!\sum\limits_{n_f}\!\left(e^{\mathcal{J}\Delta{\Psi}_{n_f,n_{\rm{RP}}}}\!-\!e^{\mathcal{J}\phi_{\rm{ant}}^{(n_r)}}\right)\right\vert^2,\!
\end{equation}where $\Delta{\Psi}$ is the residual phase after the frequency-domain calibration in \eqref{eq:NLregression} and $\mathcal{J}=\sqrt{-1}$. In the calibration, the UE is placed at different reference points (RPs), and $n_{\rm{RPs}}$ is the index of the RPs. We have used a computerized numerical control X-Y table (mm-level accuracy) to provide the ground truth of the RPs' locations. After the processing in \eqref{eq:NLregression} and \eqref{eq:ant}, we can obtain the calibration parameters ${\bf{\Upsilon}}$ and ${\phi_{\rm{ant}}^{(n_r)}}$. Note that we only need to estimate these parameters once, which can be used for future CSI calibration directly without the redundant processing again. Fig.~\ref{fig:CSI_calibration} shows an example of the phases of the CSI before and after calibration. The calibrated phases in Fig.~\ref{fig:CSI_calibration}(c) have an excellent matching compared with the real phases in Fig.~\ref{fig:CSI_calibration}(a). Even though there are several phase jumps around $-\pi$ and $\pi$ rad, the errors in the complex angle domain are small. The mean phase error is only about 0.23 rad after the CSI calibration.

\subsection{Benchmark: Active Tracking}
\label{sec:benchmark}
As mentioned in Section~\ref{sec:sys_overview}, we adopt the localization results of active tracking as the benchmark (``ground truth") for contact-free tracking. For indoor massive MIMO localization, the planar wave-front assumption is not guaranteed for the whole BS antenna array. Therefore, the super-resolution MPCs estimation algorithms, \textit{e.g.}, MUSIC, SAGE algorithm, cannot be implemented directly. In this section, we propose a tracking method exploiting the CSI signal directly. Define the ideal phase from the UE to the $n_r$-th antenna as $\phi_{n_r}=\frac{2\pi}{\lambda}\left\Vert\mathbf{P}_{n_r}\!-\!\mathbf{P}_{\rm{UE}}\right\Vert$, where $\lambda$ is the wavelength. $\mathbf{P}_{\rm{UE}}$ and $\mathbf{P}_{n_r}$ are the coordinates of the UE and the $n_r$-th antenna, respectively. Define the calibrated CSI of $n_r$-th antenna as $\tilde{H}_{n_r}$, so the corresponding phase is $\tilde{\phi}_{n_r}=\angle\tilde{H}_{n_r}$ and amplitude $\tilde{\alpha}_{n_r}=\vert\tilde{H}_{n_r}\vert$. Stemming from the concept of synthetic aperture radar (SAR), the location of the UE can be estimated via maximizing the matching function $\mathcal{C}(\mathbf{P}_{\rm{UE}})$,
\begin{equation}\label{eq:Pos_est}
\!\mathop {\arg\max}\limits_{\mathbf{P}_{\rm{UE}}}\mathcal{C}(\mathbf{P}_{\rm{UE}})=\mathop {\arg\max}\limits_{\mathbf{P}_{\rm{UE}}}\left\vert\sum\limits_{n_r=1}^{N_r}\tilde{\alpha}_{n_r}e^{\mathcal{J}\big(\phi_{n_r}-\tilde{\phi}_{n_r}\big)}\right\vert,
\end{equation}where ${N_r}=64$ is the number of antenna element at BS. For pedestrian tracking, the 2-D location is considered. So for the benchmarking, we can measure the height of the participant in advance and set it as \textit{a priori}. As $\mathcal{C}(\mathbf{P}_{\rm{UE}})$ in \eqref{eq:Pos_est} is non-convex, a common solution is conducting the grid search within the 2-D targeted area. However, the global grid search is time-consuming, especially for a large targeted area or a small grid size. To handle this problem, We propose to estimate the target's location via tracking the changes of $\mathcal{C}(\mathbf{P}_{\rm{UE}})$ based on a particle filter algorithm. For each location update, the candidate region is constrained by the pedestrian motion model, given by $\textbf{x}_{\rm{UE}}^{t+1}=\textbf{F}\textbf{x}_{\rm{UE}}^{t}+\textbf{G}n_v$, where
\begin{equation}\label{eq:MotionModel}
\textbf{x}_{\rm{UE}}=\begin{bmatrix}
\textbf{P}_{\rm{UE}}^{\top}\\ 
\textbf{v}_{\rm{UE}}^{\top}
\end{bmatrix}, \textbf{F}=\begin{bmatrix}
\mathbf{I}_{2} & \Delta t\cdot\mathbf{I}_{2}\\ 
\mathbf{0} & \mathbf{I}_{2}
\end{bmatrix}, \textbf{G}=\begin{bmatrix}
\Delta t\cdot\mathbf{1}_{2\times1}\\ 
\mathbf{1}_{2\times1}
\end{bmatrix},
\end{equation}where $\mathbf{v}_{\rm{UE}}$ is the 2-D velocity of the moving target. $\Delta t$ is the time difference between two consecutive timestamps. ${n}_v$ is the random Gaussian velocity errors with the standard deviation $\sigma_v=0.3$ m/s \cite{Chenglong2022}. In this way, we can avoid the global search for each update and speed up the computation. The $i$-th ($i=1,\cdots,K$) updated particle is weighted via $\hat{P}_w^{(i)}$, defined as \cite{Chenglong2022}
\begin{equation}\label{eq:PF_weight}
\ln\hat{P}_w^{(i)}={\frac{1}{2\sigma_c^2}\left(\min\limits_{1\leq i\leq K}\left\lbrace{P_w^{(i)}}\right\rbrace-P_w^{(i)}\right)},
\end{equation}where $P_w^{(i)}=\left(\mathcal{C}_{\rm{max}}-\mathcal{C}(\mathbf{P}^{(i)}_{\rm{UE}})\right)^2$. $K$ is the number of particles, $\sigma_c$ the standard deviation of $\mathcal{C}(\mathbf{P}_{\rm{UE}})$, and $\mathcal{C}_{\rm{max}}\!=\!1$ because we normalize $\mathcal{C}(\mathbf{P}_{\rm{UE}})$ to $(0,1]$. The normalized $\mathcal{C}(\mathbf{P}_{\rm{UE}})$ can be regarded as the target's location likelihood. 
\par

\begin{figure}[t]
\centering
\includegraphics[width=0.485\textwidth]{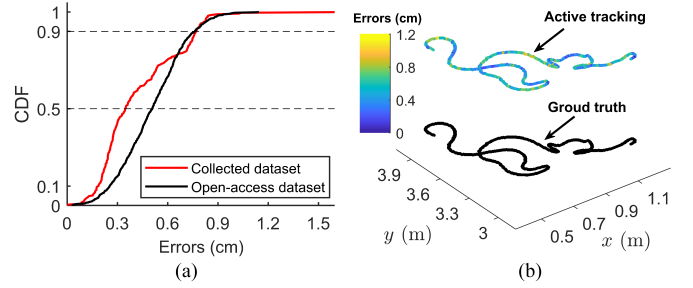}
\caption{Active tracking accuracy based on distributed massive MIMO. (a) CDF of tracking errors. (b) Tracking results of a ``cat-shape" trajectory.}
\label{fig:ActiveTrackingResults}
\end{figure}

To validate the feasibility of active tracking as the benchmark, we evaluate the localization accuracy of active tracking using the collected data based on the computerized numerical control X-Y table and our open-access distributed massive MIMO dataset \cite{CSIdatasets}. The open-access dataset was also collected via a similarly distributed antenna topology as shown in Fig.~\ref{fig:SetupPics}. In  Fig.~\ref{fig:ActiveTrackingResults}(a), the cumulative distribution functions (CDF) of the tracking errors in these two datasets are presented. Fig.~\ref{fig:ActiveTrackingResults}(b) shows an example of active tracking results, in which we formulate the tracking trajectory as a ``cat shape". We can see that the distributed massive MIMO system can guarantee millimeter-level mean tracking accuracy whereas the 90th percentile error is less than 0.78 cm. The tracking results indicate the feasibility of the proposed active tracking as the benchmark for the radar-like (contact-free) scheme. Regarding the active pedestrian tracking in the experiment, human walking is not perfectly controlled as the computerized numerical control table. But it is still reasonable to presume that the accuracy of active pedestrian tracking (\textit{e.g.}, centimeter-level is possible) is much higher than contact-free tracking in the case of distributed massive MIMO.

\section{Contact-Free Multi-Target Tracking}
\label{sec:AlgDesign}

\subsection{Target-of-Interest}
\label{sec:ToI}
In case an unknown number of pedestrians are walking in an indoor scenario, the CSI at ${n_r}$-th antenna, time $t$, and frequency $f$ (in equivalent baseband) can be given as,
\begin{equation}\label{eq:BS_CSI}
\!\!H_{n_r}(t,f)\!=\!\bar{H}_{n_r}(f)\!+\!\sum\limits_{j\in\mathcal{D}}\alpha_{n_r}^{(j)}(t,f)e^{-\mathcal{J}2\pi\left(\frac{d_{n_r}^{(j)}(t)}{\lambda}+\nu_{n_r}^{(j)}t\right)}\!,\!\!
\end{equation}where $\bar{H}_{n_r}(f)$ is the sum of all static signals. $\mathcal{D}$ denotes the set of dynamic targets. $\alpha_{n_r}^{(\cdot)}$, $d_{n_r}^{(\cdot)}$, and $\nu_{n_r}^{(\cdot)}$ denote the complex gain, the link length, and Doppler shifts at the $n_r$-th antenna element, respectively. In the framework of contact-free tracking, we are only interested in the MPCs reflected or scattered from the pedestrians. To remove the static clutters, a moving average method \cite{Sakhnini2022,Storrer2021,Berger2010} is adopted to obtain the mean CSI and then subtracted from the instantaneous CSI. So the ToI can be given as (note that we omit the indexes of $t$ and $f$ for ease of expression),
\begin{equation}\label{eq:DynCSI}
H^{\rm{ToI}}_{n_r}=\sum\limits_{j\in\mathcal{D}}\alpha_{n_r}^{(j)}e^{-\mathcal{J}\frac{2\pi}{\lambda}\left(d_{n_r}^{(j)}+\Delta d_{n_r}^{(j)}\right)}+n_{n_r},
\end{equation}where $n_{n_r}$ is the residual components after the background removal. $\Delta d_{n_r}^{(\cdot)}$ is the link length changes due to Doppler shift, given by \cite{Subedi2016}
\begin{equation}\label{eq:DopplerLen}
\Delta d_{n_r}^{(j)}=-\left(\frac{\textbf{P}_{\rm{T}}^{(j)}-\textbf{P}_{n_r}}{\Vert\textbf{P}_{\rm{T}}^{(j)}-\textbf{P}_{n_r}\Vert}+\frac{\textbf{P}_{\rm{T}}^{(j)}-\textbf{P}_{\rm{UE0}}}{\Vert\textbf{P}_{\rm{T}}^{(j)}-\textbf{P}_{\rm{UE0}}\Vert}\right)^{\top}\textbf{v}_{\rm{T}}^{(j)}t,
\end{equation}where $\textbf{P}_{\rm{T}}^{(j)}$ and $\textbf{v}_{\rm{T}}^{(j)}$ represent the location and velocity of the $j$-th target. For the pedestrians tracking, the 2-D coordinates and velocities, namely $\textbf{P}_{\rm{T}}=[x_{\rm{T}}\;y_{\rm{T}}]^{\top}$ and $\textbf{v}_{\rm{T}}=[v_{\rm{T}}^{(x)}\;v_{\rm{T}}^{(y)}]^{\top}$, are considered. Given the fact that the number of targets within the monitoring area is limited, it means that the targets' states $\textbf{x}_{\rm{T}}=[\textbf{P}_{\rm{T}}^{\top}\;\textbf{v}_{\rm{T}}^{\top}]^{\top}$ are sparsely distributed in the entire 4-D solution space $\mathbb{X}$, which reminds us of compressive sensing for location estimation. 
\par

\begin{algorithm}[t]
\label{alg:SICAR}
\caption{SICAR-based no-Doppler ToI extraction}
\DontPrintSemicolon
\KwInput{$\textbf{H}_{n_r}^{\rm{ToI}}$, $\mathbb{D}$, $\mathbb{T}$, $P_{\rm{th}}$}
\KwOutput{$\hat{H}_{n_r}^{\rm{ToI}}$}
Initialize: 
$\hat{H}_{n_r}^{\rm{ToI}}=0$;
$\textbf{X}=\textbf{H}_{n_r}^{\rm{ToI}}$;
$l=0$;\;
\While{\textsf{TRUE}}
{
	$l=l+1$;\;
	$\mathbb{G}=\mathbb{T}^{\rm{H}}\textbf{X}\mathbb{D}$;\tcp*{Range-Doppler profile}
	$(\hat{n}_{\rm{T}},\hat{n}_{\rm{D}})=\mathop {\arg\max}\limits_{n_{\rm{T}},n_{\rm{D}}}\mathbb{E}\left\{\mathbb{G}\odot\mathbb{G}^{*}\right\}$;\;
	$\phi_{n_r}^{\rm{ToI}}=\angle\mathbb{G}(\hat{n}_{\rm{T}},\hat{n}_{\rm{D}})$;\tcp*{Phase estimation}
	\tcc{Phase rotations by bandwidth \& Doppler}
	$\phi_{i}^{\rm{Range}}=2\pi(i-1)\left(\frac{\hat{n}_{\rm{T}}-1}{N_T}-0.5\right)$,$i=1,\cdots,N_f$;\;
	$\phi_{j}^{\rm{Doppler}}=2\pi(j-1)\left(\frac{\hat{n}_{\rm{D}}-1}{N_D}-0.5\right)$,$j=1,\cdots,N_t$;\;
	\tcc{Amplitude estimation}
	$\alpha_{n_r}^{\rm{ToI}}=\frac{1}{N_fN_t}\sum\limits_{i=1}^{N_f}\sum\limits_{j=1}^{N_t}e^{\mathcal{J}(\phi_{i}^{\rm{Range}}+\phi_{j}^{\rm{Doppler}})}\cdot\textbf{X}_{(i,j)}$;\;
	$\left[{\bf{\Phi}}_{n_r}^{\rm{Tmp}}\in\mathbb{C}^{N_f\times N_t}\right]_{(i,j)}=e^{-\mathcal{J}\left(\phi_{i}^{\rm{Range}}+\phi_{j}^{\rm{Doppler}}\right)}$;\;
	$\textbf{X}=\textbf{X}-\alpha_{n_r}^{\rm{ToI}}\cdot{\bf{\Phi}}_{n_r}^{\rm{Tmp}}$;\tcp*{Sequential extraction}
	$\hat{H}_{n_r}^{\rm{ToI}}=\hat{H}_{n_r}^{\rm{ToI}}+\vert\alpha_{n_r}^{\rm{ToI}}\vert e^{\mathcal{J}\phi_{n_r}^{\rm{ToI}}}$;\tcp*{Reconstruction}
	\If{$l=1$}
    {
        $\alpha_{n_r}^{\rm{max}}=\alpha_{n_r}^{\rm{ToI}}$;\tcp*{Maximal amplitude}
    }
    \If{$\vert\alpha_{n_r}^{\rm{ToI}}\vert<10^{-P_{\rm{th}}/20}\vert\alpha_{n_r}^{\rm{max}}\vert$}
    {     
        $\hat{H}_{n_r}^{\rm{ToI}}=\hat{H}_{n_r}^{\rm{ToI}}-\vert\alpha_{n_r}^{\rm{ToI}}\vert e^{\mathcal{J}\phi_{n_r}^{\rm{ToI}}}$;\;
        break;\tcp*{Delete weak component and stop}
    }
}
\textbf{Return} $\hat{H}_{n_r}^{\rm{ToI}}$.
\end{algorithm}

Define the ToIs at all of the distributed antennas as $\textbf{H}_{\rm{ToI}}=[H^{\rm{ToI}}_{1},\cdots,H^{\rm{ToI}}_{N_r}]^{\top}\in\mathbb{C}^{N_r\times1}$, so the estimation of targets' states can be done by solving
\begin{equation}\label{eq:CS_eq1}
\textbf{H}_{\rm{ToI}}=\bf{\Phi}\textbf{u}+\textbf{n},
\end{equation}where ${\bf{\Phi}}\in\mathbb{C}^{N_r\times\mathcal{M}}$ is the dictionary matrix whose $(n_r,m)$-th element is given as ${\bf{\Phi}}_{n_r,m}=e^{-\mathcal{J}\frac{2\pi}{\lambda}\left(d_{n_r}^{(m)}+\Delta d_{n_r}^{(m)}\right)}$. $\mathcal{M}=\mathcal{M}_x\mathcal{M}_y\mathcal{M}_{v_x}\mathcal{M}_{v_y}$ is the number of candidate states, where $\mathcal{M}_x$, $\mathcal{M}_y$, $\mathcal{M}_{v_x}$, and $\mathcal{M}_{v_y}$ are the number of the partitioned coordinates and velocities in $\mathbb{X}$, respectively. $\textbf{u}\in\mathbb{C}^{\mathcal{M}\times1}$ indicating the sparse solutions of targets' states, and $\left[\textbf{n}\in\mathbb{C}^{N_r\times1}\right]_{n_r}=n_{n_r}$. However, the second dimension of $\bf{\Phi}$ (\textit{i.e.}, $\mathcal{M}$) is extremely large, resulting in a long computation time. For example, suppose that the size of the monitoring area is $2$ m$\times2$ m, the walking speed ranges between $\pm1.5$ m/s, and the grid sizes of coordinates and velocities are $0.05$ m and $0.05$ m/s, respectively, then the number of candidates $\mathcal{M}=41^2\times61^2=6255001>10^6$, which will cause the computing complexity curse and is unacceptable for practical usage. However, in the framework of pedestrian tracking, we are more concerned about the coordinates and the velocities that can be obtained via a tracking filter after obtaining the locations. Furthermore, in the TDD-LTE or 5G new radio network, the downlink or uplink sub-frames are not continuously transmitted due to transmigration periodicity, which means the valid CSI sampling rate cannot be very high in practice, \textit{e.g.}, 100 Hz \cite{Daqing2022} (the setting in our experiment). In this case, the number of CSI samples $N_t$ for Doppler estimation is limited, thus the Doppler estimation accuracy is not guaranteed. By these, we can rule out the Doppler impact and only retain the location-related part in \eqref{eq:DynCSI}. Then the number of candidate states can decrease to $\mathcal{M}=\mathcal{M}_x\mathcal{M}_y=41^2=1681\ll 6255001$. 
\par
To handle this, we propose a ToI purification method based on the serial interference cancellation and reconstruction, as shown in Algorithm~\ref{alg:SICAR}. Define the ToI of the $n_r$-th antenna element at the $n_f$-th sub-carrier and $n_t$-th time slot as $\left[\textbf{H}_{n_r}^{\rm{ToI}}\in\mathbb{C}^{N_f\times N_t}\right]_{n_f,n_t}$. Define the ${N_{\rm{D}}}$-point phase-shifted discrete Fourier transform (DFT) matrix for Doppler as,
\begin{equation}
\left[\mathbb{D}\!\in\mathbb{C}^{N_t\times N_{\rm{D}}}\right]_{(n_t,n_{\rm{D}})}\!=\!\frac{1}{\sqrt{N_{\rm{D}}}}\left(e^{-\mathcal{J}\frac{2\pi}{N_{\rm{D}}}}\right)^{(n_t-1)\left(n_{\rm{D}}-1-\frac{N_{\rm{D}}}{2}\right)}\!,\nonumber
\end{equation}and the ${N_{\rm{T}}}$-point phase-shifted DFT matrix for ranging as,
\begin{equation}
\left[\mathbb{T}\!\in\mathbb{C}^{N_f\times N_{\rm{T}}}\right]_{(n_k,n_{\rm{T}})}\!=\!\frac{1}{\sqrt{N_{\rm{T}}}}\left(e^{-\mathcal{J}\frac{2\pi}{N_{\rm{T}}}}\right)^{(n_k-1)\left(n_{\rm{T}}-1-\frac{N_{\rm{T}}}{2}\right)}\!.\nonumber
\end{equation}The idea of the proposed ToI purification method is built upon removing the Doppler of each MPC via the FT sequentially, which is similar to the expectation-maximization algorithm for channel parameters estimation in \cite{Chong2002}. The iteration stops when the power of the current MPC is $P_{\rm{th}}$ dB (\textit{e.g.}, 30 dB) lower than the maximal MPC power. In this work, ${N_{\rm{D}}}$ and ${N_{\rm{T}}}$ are set as 128 and 100, respectively. Unlike many sub-space methods, the 2-D DFT in Algorithm~\ref{alg:SICAR} does not require the exact number of targets and can be implemented efficiently. According to the experimental observations, the average number of iterations using the proposed SICAR algorithm is 4.6 with rare cases having more than 8 iterations. The computing time of ToI extraction will be investigated in Section~\ref{sec:inferTime}.

\subsection{Compressive Sensing-based Positioning}
\label{sec:CBCS}
After ruling out the impact of Doppler, the dictionary matrix in \eqref{eq:CS_eq1} is simplified to ${\bf{\Phi}}_{n_r,m}=e^{-\mathcal{J}\frac{2\pi}{\lambda}d_{n_r}^{(m)}}$ which can be generated by discretizing the targeted area. The corresponding localization problem can be solved via sparsity recovery techniques, \textit{e.g.}, basic pursuit (BP), orthogonal matching pursuit (OMP), Bayesian compressive sensing (BCS), etc. Among them, BCS provides a probabilistic inference method. It efficiently solves the sparse reconstruction problem by exploiting the partial common sparse support \cite{Tipping2001}, which outperforms the conventional BP and OMP generally. However, the original BCS algorithm was designed for the real-valued problem and cannot be used directly for the complex problem in \eqref{eq:CS_eq1}. The authors in \cite{Qisong2014,Wang2022} proposed to jointly model the real and imaginary components presuming that they share the same hyper-parameters $[{\boldsymbol{\alpha}}]_m={\alpha_m}, m\in[1,\cdots,\mathcal{M}]$ (the inverse of the variances of intensity vector) and ${\beta_0}$ (the inverse of the variances of noise), and proposed complex BCS (CBCS) algorithm for the complex-value reconstruction problem. As the real and imaginary components share the same prior, the posterior density function for $\textbf{u}=[u_1,\cdots,u_{\mathcal{M}}]^{\top}$ can be given as \cite{Shihao2008,Qisong2014,Wang2022}
\begin{equation}\label{eq:CS_posterior}
\begin{aligned}
\!\!\rm{Pr}({\textbf{u}}|{\textbf{H}}_{\rm{ToI}},{\bf{\Phi}},{\boldsymbol{\alpha}},{\beta_0})&=\prod_{m=1}^{\mathcal{M}}\mathcal{N}(\mathfrak{R}(u_m)|0,\alpha_m^{-1}\beta_0^{-1})\\
&\quad\times\prod_{m=1}^{\mathcal{M}}\mathcal{N}(\mathfrak{I}(u_m)|0,\alpha_m^{-1}\beta_0^{-1})\\
&=\mathcal{CN}({\textbf{u}}|{\boldsymbol{\mu}},{\bf{\Sigma}}),
\end{aligned}
\end{equation}where
\begin{equation}\label{eq:mean_var}
\begin{aligned}{\boldsymbol{\mu}}&={\beta_0}{\bf{\Sigma}}{\bf{\Phi}}^{\rm{H}}{\textbf{H}}_{\rm{ToI}},\\
{\bf{\Sigma}}&=\left(\rm{diag}({\boldsymbol{\alpha}})+{\beta_0}{\bf{\Phi}}^{\rm{H}}{\bf{\Phi}}\right)^{-1}. 
\end{aligned}
\end{equation}
\par
Based on the CBCS algorithm, we obtain the solution $\hat{\textbf{u}}$ which has a limited number of nonzero elements. The nonzero values denote the complex amplitude and indicate the possible locations of the targets. Herein, we briefly introduce the CBCS modeling for the underlying localization problem. We refer to \cite{Shihao2009,Wang2022} for the detailed analysis and a fast-algorithm implementation which sequentially adds or deletes the candidate basis functions of the relevance vector machine (RVM) \cite{Tipping2001}. For the location estimation, instead of using the intermediate MPCs, we regard the CSI of all antennas as a whole and estimate the locations directly, so the \textbf{link association} challenge in the MPCs-based solutions can be evaded. For clarity, the essential procedures of the CBCS algorithm for instantaneous location estimates are referred to \cite{Shihao2009,Wang2022} and summarized in Algorithm~\ref{alg:CBCS}.

\begin{algorithm}[t]
\label{alg:CBCS}
\caption{CBCS algorithm for location estimates}
\DontPrintSemicolon
\KwInput{$\textbf{H}_{\rm{ToI}}$, $[{\bf{\Phi}}]_m={\bf{\Phi}}_m\in\mathbb{C}^{N_r\times1}, m\in[1,\cdots,\mathcal{M}]$}
\KwOutput{Estimated sparse vector $\hat{\textbf{u}}$}
$a=100/\rm{var}(\vert \textbf{H}_{\rm{ToI}}\vert)$; $b=1$;\;
Initialize $\alpha_m$ with a single basis vector ${\bf{\Phi}}_m$, and all other $\alpha_i\leftarrow0$, $(i\neq m)$, namely,
$s_m={\bf{\Phi}}_m^{\top}{\bf{\Phi}}_m$; $q_m={\bf{\Phi}}_m^{\top}\textbf{H}_{\rm{ToI}}$; $g_m=\textbf{H}^{\top}_{\rm{ToI}}\textbf{H}_{\rm{ToI}}+2b$; $\alpha_m=\frac{s_m(s_m-q_m^2/g_m)}{(N+2a)q_m^2/g_m+s_m}$;\;
Compute ${\boldsymbol{\mu}}$, ${\bf{\Sigma}}$, and the auxiliary variables $S_m={\bf{\Phi}}_m^{\top}{\bf{\Phi}}_m-{\bf{\Phi}}_m^{\top}{\bf{\Phi}}{\bf{\Sigma}}{\bf{\Phi}}^{\top}{\bf{\Phi}}_m$; $Q_m={\bf{\Phi}}_m^{\top}{\bf{H}}_{\rm{TOI}}-{\bf{\Phi}}_m^{\top}{\bf{\Phi}}{\bf{\Sigma}}{\bf{\Phi}}^{\top}{\bf{H}}_{\rm{TOI}}$; $G={\bf{H}}_{\rm{TOI}}^{\top}{\bf{H}}_{\rm{TOI}}-{\bf{H}}_{\rm{TOI}}^{\top}{\bf{\Phi}}{\bf{\Sigma}}{\bf{\Phi}}^{\top}{\bf{H}}_{\rm{TOI}}+2b$;\;
\While{\textsf{TRUE}}
{	
	$s_m=\frac{\alpha_mS_m}{\alpha_m-S_m}$, $q_m=\frac{\alpha_mQ_m}{\alpha_m-S_m}$, $g_m=G+\frac{Q_m^2}{\alpha_m-S_m}$;\;	
	Select a candidate basis vector ${\bf{\Phi}}_m$ and compute $\theta=\frac{(N+2a)q_m^2/g_m+s_m}{s_m(s_m-q_m^2/g_m)}$;\;
	\tcc{Refer to \cite{Shihao2009,Wang2022} for the details of sequential optimization}	
	\If{$\theta>0$ and $\alpha_m<\infty$}
	{
		Re-estimate $\alpha_m$;
	}
	\If{$\theta>0$ and $\alpha_m=\infty$}
	{
		Add ${\bf{\Phi}}_m$ to the model with updated $\alpha_m$;
	}
	\If{$\theta\leq0$ and $\alpha_m<\infty$}
	{
		Delete ${\bf{\Phi}}_m$ from the model and set $\alpha_m=\infty$;
	}
	Update ${\boldsymbol{\mu}}$, ${\bf{\Sigma}}$, $s_m$, $q_m$, and $g_m$;\;
	\If{the increase of marginal likelihood $<10^{-3}$}
	{
		Break;\tcp*{converge and terminate}
	}
}
\textbf{Return} Sparse vector $\hat{\textbf{u}}={\boldsymbol{\mu}}$.
\end{algorithm}

\subsection{Multi-Target Tracking}
\label{sec:MTT}
For contact-free MTT, it is required to associate the multiple estimates to each target at every instantaneous time slot, namely, the so-called \textbf{target and evolving association} in Section~\ref{sec:challenges}. There are several available algorithms to solve the association problem in MTT, such as JPDA \cite{Karanam2019,Storrer2021}, MHT \cite{Blackman2004}, and multi-target particle filter \cite{Sutharsan2012}. Besides the high computing complexity, they also require estimating the number of targets (or \textit{a priori}) before tracking. Alternatively, the random finite set (RFS)-based MTT has attracted great attention as it evades the explicit associations between the targeting state and the measurement state and jointly estimates the number of targets and their corresponding trajectories. 
\par
A representative RFS-based MTT algorithm is the GM-PHD filter \cite{Vo2006}. Under the assumption of independent and linear Gaussian multi-target models and Gaussian measurement models, the GM-PHD filter propagates the first moment of the multi-target posterior (\textit{a.k.a}, the intensity function). It provides a computationally efficient closed-form alternative to the conventional PHD filter. GM-PHD filter is reasonable for our MTT objective because the sparse state in \eqref{eq:CS_posterior} of the CBCS algorithm has already been modeled as a Gaussian distribution and each pedestrian moves independently which is also indicated by the signal model in \eqref{eq:DynCSI}. Without considering the spawned targets, the targets' states (\textit{i.e.}, coordinates and velocities) at time $t$ can be modeled by the union of the surviving targets from previous time $t-1$ and the spontaneous births at time $t$. The predicted intensity $u_{t|t-1}$ associated with the multi-target predicted density can be given by \cite{Vo2006}, 
\begin{equation}\label{eq:state_predict}
u_{t|t-1}(\textbf{x}_{\rm{T}})=u_{S,t|t-1}(\textbf{x}_{\rm{T}})+\gamma_t(\textbf{x}_{\rm{T}}),
\end{equation}where\addtocounter{equation}{-1}
\begin{subequations}
\begin{align}
\!\!\!u_{S,t|t-1}(\textbf{x}_{\rm{T}})&=p_S\sum\limits_{i=1}^{J_{S,t-1}}\omega_{t-1}^{(i)}\mathcal{N}(\textbf{x}_{\rm{T}};\textbf{m}_{S,t|t-1}^{(i)},\textbf{P}_{S,t|t-1}^{(i)}),\label{eq:state_predict_uS}\\
\!\!\!\gamma_t(\textbf{x}_{\rm{T}})&=\sum\limits_{i=1}^{J_{\gamma,t}}\omega_{t}^{(i)}\mathcal{N}(\textbf{x}_{\rm{T}};\textbf{m}_{t}^{(i)},\textbf{P}_{t}^{(i)})\label{eq:state_predict_gamma},
\end{align}
\end{subequations}where $p_S$ is the target survival probability. $J_{(\cdot)}$ represents the number of Gaussian mixtures. $\omega_{(\cdot)}$ is the weight of Gaussian intensity. For contact-free tracking, we also adopt the constant velocity model as in active tracking in Section~\ref{sec:benchmark}. So we have the mean state update as $\textbf{m}_{S,t|t-1}^{(i)}=\textbf{F}\textbf{m}_{t-1}^{(i)}$ and the covariance update as $\textbf{P}_{S,t|t-1}^{(i)}=\sigma_{v}^2\textbf{G}\textbf{G}^{\top}+\textbf{F}\textbf{P}_{t-1}^{(i)}\textbf{F}^{\top}$, where $\sigma_{v}$, $\textbf{F}$, and $\textbf{G}$ define the pedestrian motion model as in \eqref{eq:MotionModel}. In \eqref{eq:state_predict_gamma}, the intensity of the new-born targets $\gamma_t(\textbf{x}_{\rm{T}})$ is also assumed as a Gaussian mixture \cite{Vo2006}. To handle the case that the newly emerging targets may appear anywhere and anytime, generally, the newborn targets are initialized to cover the entire target area \cite{Ristic2012}. However, it is not necessary for our MTT framework as we use the instantaneous locations estimated by the CBCS algorithm which covers the entire monitoring region naturally.
\par

\begin{algorithm}[t]
\label{alg:GM-PHD}
\caption{GM-PHD filter for multi-target tracking}
\DontPrintSemicolon
\KwInput{$\textbf{Z}_t$, $P_S$, $\textbf{F}$, $\sigma_{v}$, $\textbf{G}$, $\omega_{\rm{min}}$, $D_{\rm{min}}$, $J_{\rm{max}}$}
\KwOutput{Multi-target states $\hat{\textbf{X}}_t$}
$j=0$;\;
\ForEach {new-born target $i\in\{1,\cdots,J_{\gamma,t}\}$}
{
	$j\leftarrow j+1$;\;
	\tcc{Initialize}
	$\omega_{t|t-1}^{(j)}=\omega_{\gamma|t}^{(i)}$;\;
	$\textbf{m}_{t|t-1}^{(j)}=\textbf{m}_{\gamma|t}^{(i)}$;\;
	$\textbf{P}_{t|t-1}^{(j)}=\textbf{P}_{\gamma|t}^{(i)}$;\;
}
\ForEach {surviving target $i\in\{1,\cdots,J_{S,t-1}\}$}
{
	$j\leftarrow j+1$;\tcp*{Combine with new-born targets}
	\tcc{State prediction}
	$\omega_{t|t-1}^{(j)}=P_{S}\omega_{\gamma|t-1}^{(i)}$;\;
	$\textbf{m}_{t|t-1}^{(j)}=\textbf{F}\textbf{m}_{\gamma|t-1}^{(i)}$;\;
	$\textbf{P}_{t|t-1}^{(j)}=\sigma_{v}^2\textbf{G}\textbf{G}^{\top}+\textbf{F}\textbf{P}_{t-1}^{(i)})\textbf{F}^{\top}$;\;
}
$J_{t|t-1}=J_{\gamma,t}+J_{S,t-1}$;\tcp*{Number of components}
$l=0$;\;
\ForEach {$z\in\textbf{Z}_t$}
{
	$l\leftarrow l+1$;\;
	\ForEach {$i=1,\cdots,J_{t|t-1}$}
	{
		Update target states via \eqref{eq:State_update_ut}-\eqref{eq:State_update_qt};\;
		\textbf{Return} $\omega_{t}^{(l\cdot J_{t|t-1}+i)},\textbf{m}_{t}^{(l\cdot J_{t|t-1}+i)},\textbf{P}_{t}^{(l\cdot J_{t|t-1}+i)}$;\;
	}
}
{Pruning, merging, and capping based on the predefined threshold $\omega_{\rm{min}},D_{\rm{min}},J_{\rm{max}}$;\tcp*{refer to \cite{Vo2006}}
\textbf{Return} Pruned components $\hat{\omega}_{t},\hat{\textbf{m}}_{t},\hat{\textbf{P}}_{t}$;\;}
$\hat{\textbf{X}}_t=\emptyset$;\;
\ForEach {pruned component $\hat{\omega}_{t}>0.5$}
{
	$\hat{\textbf{X}}_t=\begin{bmatrix}
	\hat{\textbf{X}}_t&\hat{\textbf{m}}_{t}
	\end{bmatrix}$;\tcp*{Multi-target state saving}
}
\textbf{Return} Multi-target states $\hat{\textbf{X}}_t$.
\end{algorithm}

The measurement states $Z_t$ are the instantaneous location estimates at time $t$ based on the CBCS algorithm in Section~\ref{sec:CBCS}. The estimates include the locations of the real targets and some false alarms. The posterior intensity under a Gaussian mixture model can be given by \cite{Vo2006},
\begin{equation}\label{eq:Obs_update}
u_{t}(\textbf{x}_{\rm{T}})=(1-p_D)u_{t|t-1}(\textbf{x}_{\rm{T}})+\sum\limits_{\textbf{z}\in Z_{t}}u_{D,t}(\textbf{x}_{\rm{T}};\textbf{z}),
\end{equation}where $p_D=0.95$ is the detection probability. The target states are updated via the new CBCS estimates within the framework of Kalman recursion \cite{Vo2006},\addtocounter{equation}{-1}
\begin{subequations}
\begin{align}
u_{D,t}(\textbf{x}_{\rm{T}};\textbf{z})&=\sum\limits_{i=1}^{J_{t|t-1}}\omega_{t}^{(i)}(\textbf{z})\mathcal{N}\left(\textbf{x}_{\rm{T}};\textbf{m}_{t|t}^{(i)}(\textbf{z}),\textbf{P}_{t|t}^{(i)}\right),\label{eq:State_update_ut}\\
\omega_{t}^{(i)}(\textbf{z})&=\frac{p_D\omega_{t|t}^{(i)}q_t^{(i)}(\textbf{z})}{\kappa_t(\textbf{z})+p_D\sum\limits_{j=1}^{J_{t|t-1}}\omega_{t|t}^{(j)}q_t^{(j)}(\textbf{z})},\label{eq:State_update_omega}\\
\textbf{m}_{t|t}^{(i)}(\textbf{z})&=\textbf{m}_{t|t-1}^{(i)}+\textbf{K}_t^{(i)}(\textbf{z}-\textbf{H}_t\textbf{m}_{t|t-1}^{(i)}),\label{eq:State_update_m}\\
\textbf{P}_{t|t}^{(i)}&=\left[\textbf{I}-\textbf{K}_t^{(i)}\textbf{H}_t\right]\textbf{P}_{t|t-1}^{(i)},\label{eq:State_update_P}\\
\textbf{K}_t^{(i)}&=\textbf{P}_{t|t-1}^{(i)}\textbf{H}_t^{\top}\left(\textbf{H}_t\textbf{P}_{t|t-1}^{(i)}\textbf{H}_t^{\top}+\textbf{R}_t\right)^{-1},\label{eq:State_update_K}\\
 q_t^{(j)}(\textbf{z})&=\mathcal{N}\left(\textbf{z};\textbf{H}_t\textbf{m}_{t|t-1}^{(i)},\textbf{R}_t+\textbf{H}_t\textbf{P}_{t|t-1}^{(i)}\textbf{H}_t^{\top}\right)\label{eq:State_update_qt}.
\end{align}
\end{subequations}where $\kappa_t(\textbf{z})$ represents the clutter intensity which is generally assumed to follow a uniform distribution within the target space. $\textbf{H}_t=[\textbf{I}_2\;\textbf{0}_2]$ because we only have the location estimates from the CBCS algorithm for updating. $\textbf{R}_t$ is the observation noise covariance matrix. Moreover, in the context of the GM-PHD filter, pruning, merging, and capping are usually adopted to relieve the computing pressure. Pruning is to truncate the components having weaker weights than the defined threshold which is set as $\omega_{\rm{min}}=10^{-5}$ \cite{Vo2006}. Merging is to combine the closely-located components. In our work, we merge the components with a distance smaller than $D_{\rm{min}}=0.2$ m. Capping is to disregard the weakest components when the number of components is larger than the given maximal number of Gaussian components, which is set as $J_{\rm{max}}=50$ according to \cite{Vo2006}. For clarity, the essential procedures of the GM-PHD filter for multi-target tracking are referred to \cite{Vo2006} and summarized in Algorithm~\ref{alg:GM-PHD}.

\begin{figure}[t]
\centering
\includegraphics[width=0.49\textwidth]{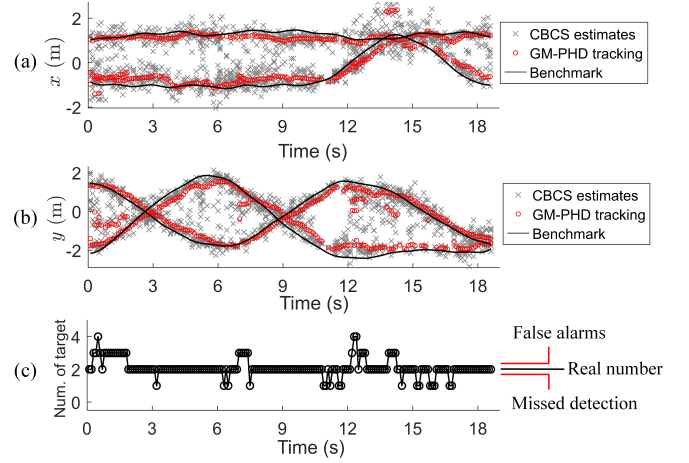}
\caption{CBCS positioning and the GM-PHD tracking results versus benchmark: (a)-(b) $x$ and $y$ coordinates. (c) Estimates of the number of targets.}
\label{fig:xyTracking}
\end{figure}

\begin{figure*}[t]
\centering
\setlength{\abovecaptionskip}{-0.05cm}
\includegraphics[width=0.99\textwidth]{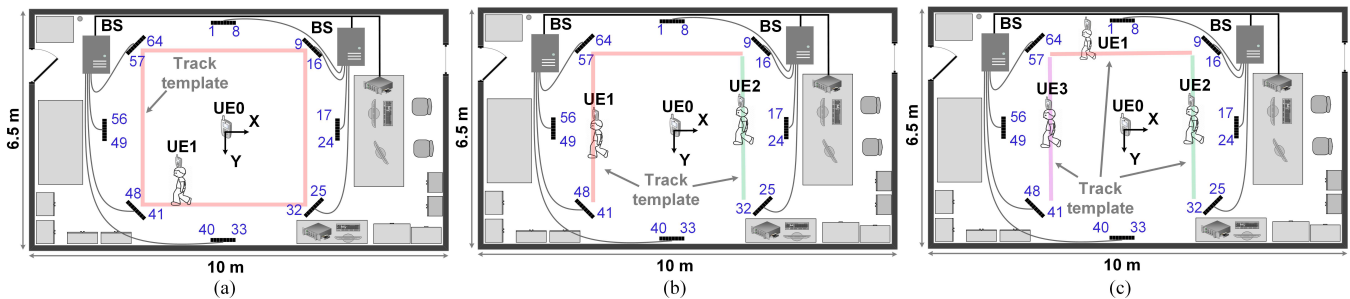}
\caption{Room layout and the setup deployment for the contact-free tracking in cases of (a) a single person, (b) two persons, and (c) three persons.}
\label{fig:FloorPlan}
\end{figure*}

\begin{algorithm}[t]
\label{alg:CSI-MTT}
\caption{The proposed contact-free MTT algorithm for the distributed massive MIMO system}
\DontPrintSemicolon
\ForEach {new CSI measurement}
{
	CSI calibration using ${\bf{\Upsilon}}$ and ${\phi_{\rm{ant}}^{(n_r)}}$, $n_r=1,\cdots,N_r$;\;
	Background removal;\;
	\ForEach {CSI data in 0.1-s time interval}
	{
		\For{$n_r=1:N_r$}
		{
			ToI extraction based on Algorithm~\ref{alg:SICAR};\;
			\textbf{Return} $\hat{H}_{n_r}^{\rm{ToI}}$;\;
		}
		CBCS for location estimates via Algorithm~\ref{alg:CBCS};\;
		GM-PHD for tracking via Algorithm~\ref{alg:GM-PHD};\;
		\textbf{Return} Updated states of the targets.\;
	}
}
\end{algorithm}

\subsection{Putting Things Together}
\label{sec:PutTogether}
This subsection summarizes the proposed contact-free MTT algorithm for the radar-like massive MIMO prototype. When the newly measured CSI is received, we conduct the calibration first using the parameters obtained in Section~\ref{sec:calibration}. Then the background static components are removed by subtracting the moving average CSI. In this work, we set the location update time window as 0.1 s, which is also widely adopted for pedestrian tracking \cite{widar2,Zhongqin2022}. Within the time window, we presume the changes in pedestrians' location are negligible and the Doppler velocity is constant. As the sampling rate of the established TDD massive MIMO system is 100 Hz, we conduct ToI extraction and MTT using 10 CSI data. In ToI extraction, Algorithm~\ref{alg:SICAR} is conducted for each antenna. Then, the ToI of all antennas is regarded as a whole and fed to the CBCS algorithm. GM-PHD filter will track the motion of the multi-target based on the instantaneous location estimates from CBCS. The number of predicted states in the GM-PHD filter indicates the number of targets \cite{Vo2006}. The procedures of the proposed contact-free MTT algorithm are given in Algorithm~\ref{alg:CSI-MTT}. 
\par
Fig.~\ref{fig:xyTracking} shows an example of the CBCS instantaneous location estimates and the GM-PHD filter-based tracking results for two-pedestrian tracking. In Figs.~\ref{fig:xyTracking}(a)-(b), we can see that the CBCS algorithm estimates the targets' locations correctly yet messily. The red circles represent the tracking results after feeding to the GM-PHD filter, which shows a smoother trajectory with a limited number of outliers. Fig.~\ref{fig:xyTracking}(c) shows the predicted number of targets based on the GM-PHD filter. In most cases, it obtains correct estimates but has some false alarms corresponding to the short lifetime trajectories and missed detection. This problem can be solved by jointly propagating the posterior intensity and the posterior cardinality distribution, such as the cardinalized PHD filter \cite{Vo2007}. We would like to leave its adaption to our distributed massive MIMO radar-like system for future exploration.

\section{Performance Evaluation}
\label{sec:PerfEva}
In this section, we evaluate the performance of contact-free human tracking using the sub-6 GHz distributed massive MIMO communication testbed. Three participants have been invited for the single-person tracking and multi-person tracking. Fig.~\ref{fig:FloorPlan} shows the room layout of the measurement campaign and the setup deployment for the three cases: single-person tracking, two-person tracking, and three-person tracking, respectively. For each scenario, the participant walks along the predefined track templates. As mentioned in Section~\ref{sec:campaign}, we ask for each participant to carry an additional transmitter (\textit{i.e.}, UE1, UE2, and UE3 in Fig.~\ref{fig:FloorPlan}) for benchmarking. The performance and the feasibility of active tracking have been validated in Section~\ref{sec:benchmark}. For the tracking performance evaluation, we calculate the distance between the locations of contact-free tracking and benchmark at each time instant. 

\subsection{Single-Pedestrian Tracking}
\label{sec:SingleEva}
\subsubsection{Accuracy Evaluation}
\label{sec:SingleAccEva}
\begin{figure}[t]
\centering
\setlength{\abovecaptionskip}{-0.05cm}
\includegraphics[width=0.485\textwidth]{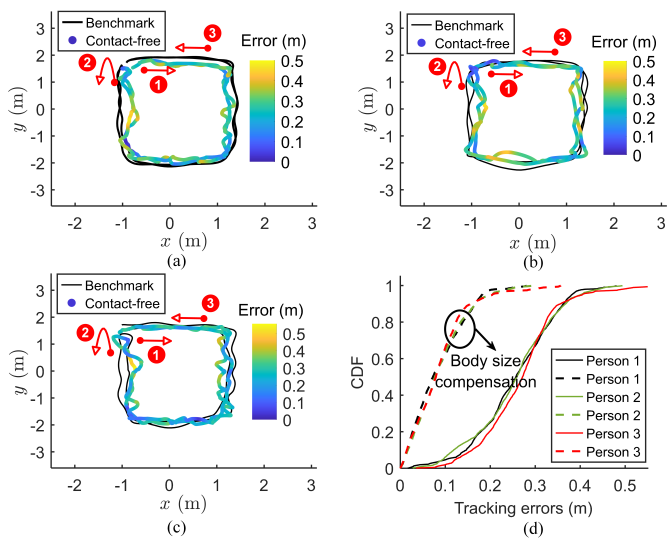}
\caption{Contact-free single-person tracking results: (a)-(c) Tracking trajectories versus benchmark for three individuals (arrows indicate the walking directions). (d) CDF of tracking errors (dash lines denote the errors after body size compensation under the cylindrical body hypothesis).}
\label{fig:SinglePersonTracking}
\end{figure}

Fig.~\ref{fig:SinglePersonTracking} shows the contact-free single-person tracking results using the proposed algorithm, in which the dot color depth indicates the tracking errors along the trajectory. In this scenario, each individual walks along a rectangular track as presented in Fig.~\ref{fig:FloorPlan}(a). After reaching the start location for the first time, the person takes a U-turn and backtracks. The red arrows in Figs. \ref{fig:SinglePersonTracking}(a)-(c) illustrate the walking directions. Fig.~\ref{fig:SinglePersonTracking}(d) summarizes the contact-free tracking errors for the three participants. The proposed algorithm achieves decimeter-level accuracy with the largest errors around 50 cm. Importantly, we can distinctly observe that the contact-free tracking trajectories are smaller rectangles compared with the benchmark results, which are closer to the transmitter UE0. This phenomenon can be explained by the truth that the human reflector is not a single point. In practice, the signal transmitted by the UE0 will be reflected/scattered by the curved body surface (torso or limbs) facing the UE0. This causes the reflection points to have an offset to the center of the human body, which roughly equals the radius of the torso (15 to 30 cm generally) under the cylindrical body hypothesis \cite{Ghaddar2007}. In our experiment, the antenna for benchmarking is placed on top of the head, which obtains the locations of the body centroid during movement. So the compensated tracking errors are calculated based on the distance from the estimates to the periphery of the cylindrical body given by $\Vert \hat{\textbf{P}_{\rm{T}}}-\textbf{P}\Vert$, where $\textbf{P}$ is the cylindrical periphery, given by $\lbrace\textbf{P}|\Vert\textbf{P}-\textbf{P}_{\rm{BM}}\Vert=r_{\rm{h}}\rbrace$, where $\textbf{P}_{\rm{BM}}$ is the location of the benchmark. $r_{\rm{h}}$ is the radius of the torso which is set as 20 cm in this work. The contact-free tracking accuracy after the body size compensation is presented by the dashed plots in Fig.~\ref{fig:SinglePersonTracking}(d). The median error is about 7.7 cm whereas the 95th percentile error is 16.9 cm. It should be noted that we adopt the tracking accuracy after the body size compensation for the performance evaluation hereinafter.

\subsubsection{Impact of Antenna Number}
\label{sec:SingleAntNum}
\begin{figure}[t]
\centering
\setlength{\abovecaptionskip}{-0.05cm}
\includegraphics[width=0.485\textwidth]{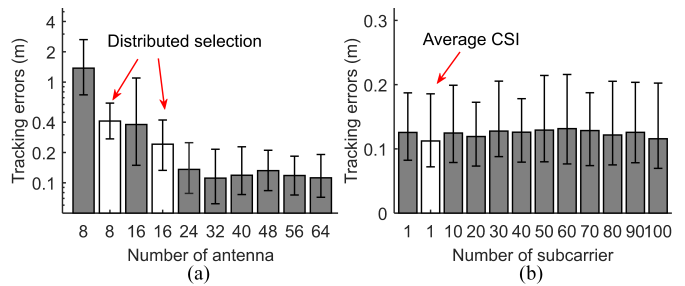}
\caption{Massive MIMO radar-like 75th percentile tracking errors with low bound 50th percentile errors and upper bound 95th percentile errors: (a) Impact of the number of antennas. (b) Impact of the number of sub-carriers.}
\label{fig:AntFreNumEva}
\end{figure}

Moreover, we evaluate the contact-free tracking performance concerning the impact of the number of antennas and the bandwidth, as shown in Fig.~\ref{fig:AntFreNumEva}. For the impact of the antenna number evaluation, we first select the antenna elements sequentially as the labeled order shown in Fig.~\ref{fig:FloorPlan}. For example, we use the sequential antenna elements from $\#1$ to $\#24$ for the case of 24-antenna evaluation. Fig.~\ref{fig:AntFreNumEva}(b) presents the tracking errors of the varied antenna elements via the 75th percentile error bars with the low bound 50th percentile and upper bound 95th percentile errors, respectively. Increasing the number of antenna elements improves the tracking accuracy and reduces the outliers generally. Nevertheless, the accuracy saturates both in accuracy and precision when the number of antennas reaches 24, in which the 75th percentile errors are less than 13.6 cm and 95th percentile errors are less than 25.1 cm. 
\par
Furthermore, when considering the antenna deployment topology, a better tracking performance can be expected, especially for the cases of fewer antenna elements, \textit{e.g.}, 8 or 16. The proposed algorithm does not use the AoA but the CSI directly for positioning and tracking, so the antenna array construction is not necessary. For the 8-antenna  evaluation, we select the antenna elements $\{\#4, \#5, \#20, \#21, \#36, \#37, \#52, \#53\}$ labeled in Fig.~\ref{fig:FloorPlan}. For the 16-antenna evaluation, besides the above selected 8 antennas, we include another 8 antenna elements $\{\#12, \#13, \#28, \#29, \#44, \#45, \#60, \#61\}$. Note that it cannot be expected to list all the possible selections. These two typical cases allowing antennas to be distributed as uniformly as possible are considered. As shown in \ref{fig:AntFreNumEva}(a), in the case of 8 antennas, the 75th and 95th percentile tracking errors using the distributed selection decrease from 1.38 m and 2.64 m to 0.41 m and 0.62 m, respectively. In the case of 16 antennas, the 75th and 95th percentile errors decrease by 0.14 m and 0.68 m, respectively. The boosting performance can be explained by the geometrical diversity of the distributed antenna selection. It also indicates that antenna placement optimization is essential for practical implementation, which deserves further investigation and we would like to leave it for future work.

\subsubsection{Impact of Bandwidth}
\label{sec:SingleBW}
In Fig.~\ref{fig:AntFreNumEva}(b), the impact of bandwidth or number of sub-carrier on tracking performance is presented. We can observe that when we decrease the number of adopted sub-carriers, the tracking performance has no clear degradation even if a single sub-carrier is adopted. The fluctuations of the 75th and 95th percentile errors are less than 1.8 cm and 4.3 cm, respectively. This is reasonable because the adopted massive MIMO communication system is with a standard cellular bandwidth of 18 MHz. The sub-carrier interval is 180 kHz for 100 sub-carriers. The limited bandwidth causes a coarse-grained ranging resolution. On the other hand, due to the limited contribution of bandwidth to tracking performance, the Fourier transform for ranging in Algorithm~\ref{alg:SICAR} can be disregarded to reduce the computing complexity. For example, we can average the CSI along the frequency domain before ToI extraction to speed up the location inference. The tracking accuracy using the average CSI is shown in Fig.~\ref{fig:AntFreNumEva}(b).

\subsection{Multiple Pedestrians Tracking}
\begin{figure}[t]
\centering
\setlength{\abovecaptionskip}{-0.05cm}
\includegraphics[width=0.485\textwidth]{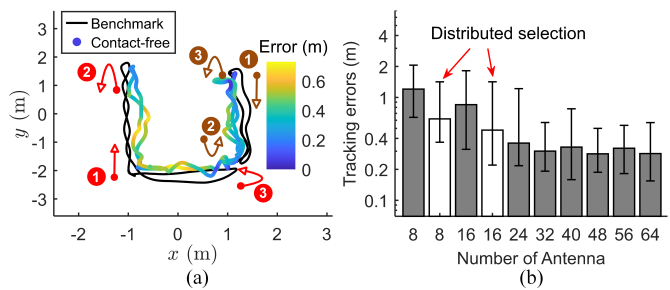}
\caption{Contact-free two-person tracking results: (a) Tracking trajectories versus benchmark (arrows with different colors indicate the walking directions of different pedestrians). (b) Impact of antenna numbers on tracking accuracy.}
\label{fig:TwoPersonEva}
\end{figure}

\begin{figure}[t]
\centering
\setlength{\abovecaptionskip}{-0.05cm}
\includegraphics[width=0.485\textwidth]{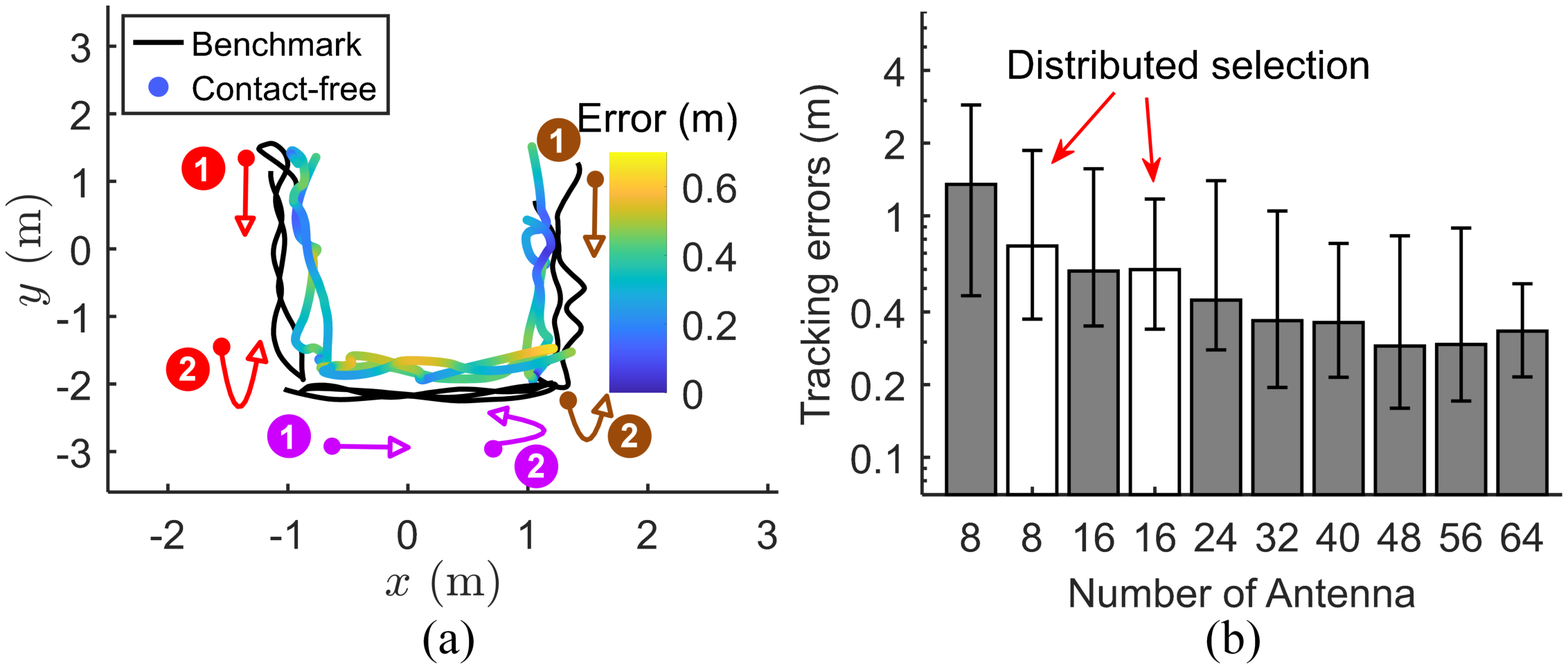}
\caption{Contact-free three-person tracking results: (a) Tracking trajectories versus benchmark (arrows with different colors indicate the walking directions of different pedestrians). (b) Impact of antenna numbers on tracking accuracy.}
\label{fig:ThreePersonEva}
\end{figure}

The measurement campaigns for two-person and three-person tracking are given in Figs.~\ref{fig:FloorPlan}(b)-(c). In Fig.~\ref{fig:TwoPersonEva}(a) and Fig.~\ref{fig:ThreePersonEva}(a), the tracking results of two pedestrians and three pedestrians versus the benchmarks are presented, in which all 64 antennas are adopted for the tracking. The arrows with different colors represent the walking directions of different targets. Similar to the case of single-person tracking, the estimated trajectories are closer to the transmitter UE0 due to the reflections/scattering on the periphery of the torsos. Namely, body size compensation is adopted for the tracking accuracy evaluation. Besides, we observe that the tracking errors of the multi-pedestrian cases are larger than the single-pedestrian tracking. In Table~\ref{table:MTT_PerErr}, we summarize the percentile errors concerning different numbers of targets. The median error for single-person tracking is only 7.3 cm and the 95th percentile error is 18.2 cm. But when the number of targets is three, the median error increases to 19.1 cm and the 95th percentile error to 42.2 cm. The possible reasons are two-fold: First, when there are multiple targets, the signal reflected by one target may be partially blocked by another person. This results in the reflected signal cannot be detected by the receivers behind. This is equivalent to the case of fewer adopted antennas. Second, we use DFT for the ToI extraction in Section~\ref{sec:ToI}, which may have worse Doppler resolution for multiple targets due to the side-lobe power leakage of DFT. Moreover, the low sampling rate of our TDD-OFDM system also challenges the ToI extraction. In other words, the signal-to-noise ratio (SNR) of the ToI will decrease, which causes tracking performance degradation.
\par

\begin{table}[t]
\centering
\caption{Percentile tracking errors of the proposed contact-free MTT algorithm in meter}
\begin{tabular}{l|c|c|c}
\hline
                           & \textbf{50th}    & \textbf{75th}    & \textbf{95th} \\ \hline\hline
\textbf{Single pedestrian} & 0.073            & 0.127            & 0.182         \\ \hline
\textbf{Two pedestrians}   & 0.148            & 0.254            & 0.457         \\ \hline
\textbf{Three pedestrians} & 0.191            & 0.289            & 0.422         \\ \hline
\end{tabular}
\label{table:MTT_PerErr}
\end{table}

Moreover, the impact of antenna number on MTT accuracy is also investigated in Fig.~\ref{fig:TwoPersonEva}(b) and Fig.~\ref{fig:ThreePersonEva}(b). Note that the impact of bandwidth is not analyzed here because of its little influence on tracking accuracy as presented in Section~\ref{sec:SingleBW}. The average CSI along sub-carriers is used for the MTT. The tracking accuracy improves as more antennas are utilized. However, the outliers in multi-pedestrian tracking increase distinctly compared with the case of a single pedestrian, which are indicated by the larger offsets from 75th to 95th percentile errors. For the smaller number of antennas, \textit{i.e.}, 8 and 16, we consider the distributed antenna selections as in Section~\ref{sec:SingleAntNum}. An accuracy improvement is observed more or less, but the overall performance is not promising concerning the size of the monitoring area in the experiment, because the 75th and 95th percentile errors are larger than 0.59 m and 1.17 m, respectively. These observations indicate that a large number of antennas is necessary for the bandwidth-limited system to guarantee a satisfying contact-free MTT accuracy.

\begin{figure}[t]
\centering
\setlength{\abovecaptionskip}{-0.05cm}
\includegraphics[width=0.485\textwidth]{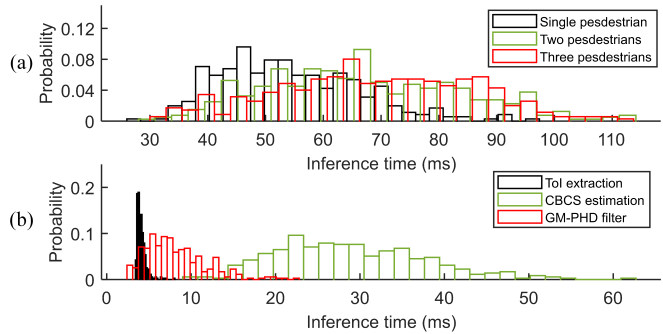}
\caption{Histograms of inference time for each location updating: (a) Different number of pedestrians. (b) Time consumption for the ToI extraction, CBCS algorithm, and GM-PHD filter in case of a single pedestrian.}
\label{fig:TimeHistogram}
\end{figure}

\subsection{Inference Time}
\label{sec:inferTime}
In this work, we adopt a 0.1-s time interval for CSI data accumulation and location updating, which means the real-time tracking is guaranteed if the location updating latency is less than 0.1 s. The running environment for evaluation is Dell OptiPlex 7050 with an Intel(R) Core(TM) i7-7700 CPU@3.60GHz and 16 GB RAM. The latency is calculated by the \texttt{tic-toc} commands in MATLAB 2021b. Fig.~\ref{fig:TimeHistogram}(a) shows the histograms of the inference time for different numbers of targets, in which the inference latency is less than 0.1 s mostly. The single-person tracking has the fastest updating in general with a 54.2-ms mean latency, whereas two- and three-person tracking has similar inference latency with mean values of 66.8 ms and 68.9 ms, respectively. These results testify the proposed contact-free tracking algorithm can achieve real-time processing for MTT purposes. We further evaluate the inference latency for the major algorithm blocks, namely, ToI extraction, CBCS location estimation, and GM-PHD filter, as shown in Fig.~\ref{fig:TimeHistogram}(b). Distinctly, the CBCS algorithm has the largest time consumption with a 28.9-ms mean value. This also shows the significance of the ToI purification in Section~\ref{sec:ToI} to avoid the computing complexity curse.

\section{Conclusion}
\label{sec:conclusion}
In this paper, a radar-like distributed massive MIMO prototype has been built upon an available sub-6 GHz OFDM communication system. Based on the established prototype, we investigated the contact-free multi-pedestrian tracking problems. Instead of using the intermediate geometrical features (AoA, ToF, Doppler, etc.), we proposed to localize and track the multiple targets exploiting the CSI signal directly. Specifically, a recursive algorithm combining the CBCS and the GM-PHD filter was proposed for multiple pedestrians tracking in real time. According to the experimental results, we achieved the 75th and 95th percentile tracking accuracy of 12.7 cm and 18.2 cm for single-pedestrian tracking and 28.9 cm and 45.7 cm for multi-pedestrian tracking, respectively. 
\par
This paper demonstrates our preliminary investigation of the sub-6 GHz distributed massive MIMO system for its radar-like functionality. It deserves further and extensive exploration. The future work will consist of both experimental analysis and algorithm development. For future experiments, a larger number of targets in various indoor scenarios will be investigated. The placement optimization for the massive antenna array in a specific scenario will be considered for the performance enhancement. Besides moving targets, how to localize multiple static targets and hybrid dynamic and static targets will be considered.

\bibliographystyle{IEEEtran}
\bibliography{MaMIMO_radar}

\end{document}